\documentclass[a4paper,conference]{IEEEtran}

\usepackage{color,soul} 

\usepackage{booktabs}
\usepackage{multirow}
\usepackage[hidelinks]{hyperref}

\usepackage{siunitx}
\usepackage{url}
\usepackage{makecell}

\usepackage{blindtext}

\usepackage{algorithm}
\usepackage{algpseudocode}

\usepackage{comment}

\usepackage{graphicx} 

\usepackage{amssymb}
\usepackage{listings} 
\usepackage{verbatim}
\usepackage{booktabs}
\usepackage{float}
\usepackage{tablefootnote}

\usepackage{tabularx} 
\usepackage{xcolor}

\usepackage{amsmath}
\usepackage[nameinlink,noabbrev]{cleveref}
\usepackage{amsmath,amssymb,amsfonts}
\usepackage{graphicx}
\usepackage{textcomp}
\usepackage{listings} 
\usepackage{mdframed}

\IEEEoverridecommandlockouts
\usepackage{cite}
\usepackage{amsmath,amssymb,amsfonts}
\usepackage{graphicx}
\usepackage{textcomp}
\usepackage{xcolor}
\def\BibTeX{{\rm B\kern-.05em{\sc i\kern-.025em b}\kern-.08em
    T\kern-.1667em\lower.7ex\hbox{E}\kern-.125emX}}
\begin{document}

\title{Attack on a PUF-based Secure \\Binary Neural Network}





    
\author{
    \IEEEauthorblockN{Bijeet Basak, Nupur Patil, Kurian Polachan, Srinivas Vivek}
    \IEEEauthorblockA{International Institute of Information Technology, Bangalore, India\\
    Email: \{bijeet.basak, nupur.patil, kurian.polachan, srinivas.vivek\}@iiitb.ac.in}
}


\maketitle

\begin{abstract}
Binarized Neural Networks (BNNs) deployed on memristive crossbar arrays provide energy-efficient solutions for edge computing but are susceptible to physical attacks due to memristor nonvolatility. Recently, Rajendran et al. (IEEE Embedded Systems Letter 2025) proposed a Physical Unclonable Function (PUF)-based scheme 
to secure BNNs against theft attacks. Specifically, the  weight and bias matrices of the BNN layers were secured by swapping columns based on device's PUF key bits.

In this paper, we demonstrate that this scheme to secure BNNs is vulnerable to PUF-key recovery attack. As a consequence of our attack, we recover the secret weight and bias matrices of the BNN. Our approach is motivated by differential cryptanalysis and reconstructs the PUF key bit-by-bit by observing the change in model accuracy, and eventually recovering the BNN model parameters. Evaluated on a BNN trained on the MNIST dataset, our attack could recover 85\% of the PUF key, and recover the BNN model up to 93\% classification accuracy compared to the original model's 96\% accuracy. Our attack is very efficient and it takes a couple of minutes to recovery the PUF key and the model parameters.

\end{abstract}


\section{Introduction}

Deep Neural Networks (DNNs) have become an important foundation of modern artificial intelligence, enabling breakthroughs in diverse fields such as medical imaging, natural language processing, autonomous driving, and robotics. These networks consist of multiple layers of interconnected nodes that learn complex representations of data through large-scale training on labeled datasets. Despite their success, conventional DNNs require significant computational and memory resources, often necessitating powerful processors and substantial energy. This makes them challenging to deploy on low-power devices such as smartphones, IoT nodes, or embedded systems.

\begin{figure*}[!htbp]
    \centering
    \includegraphics[width=0.99\linewidth]{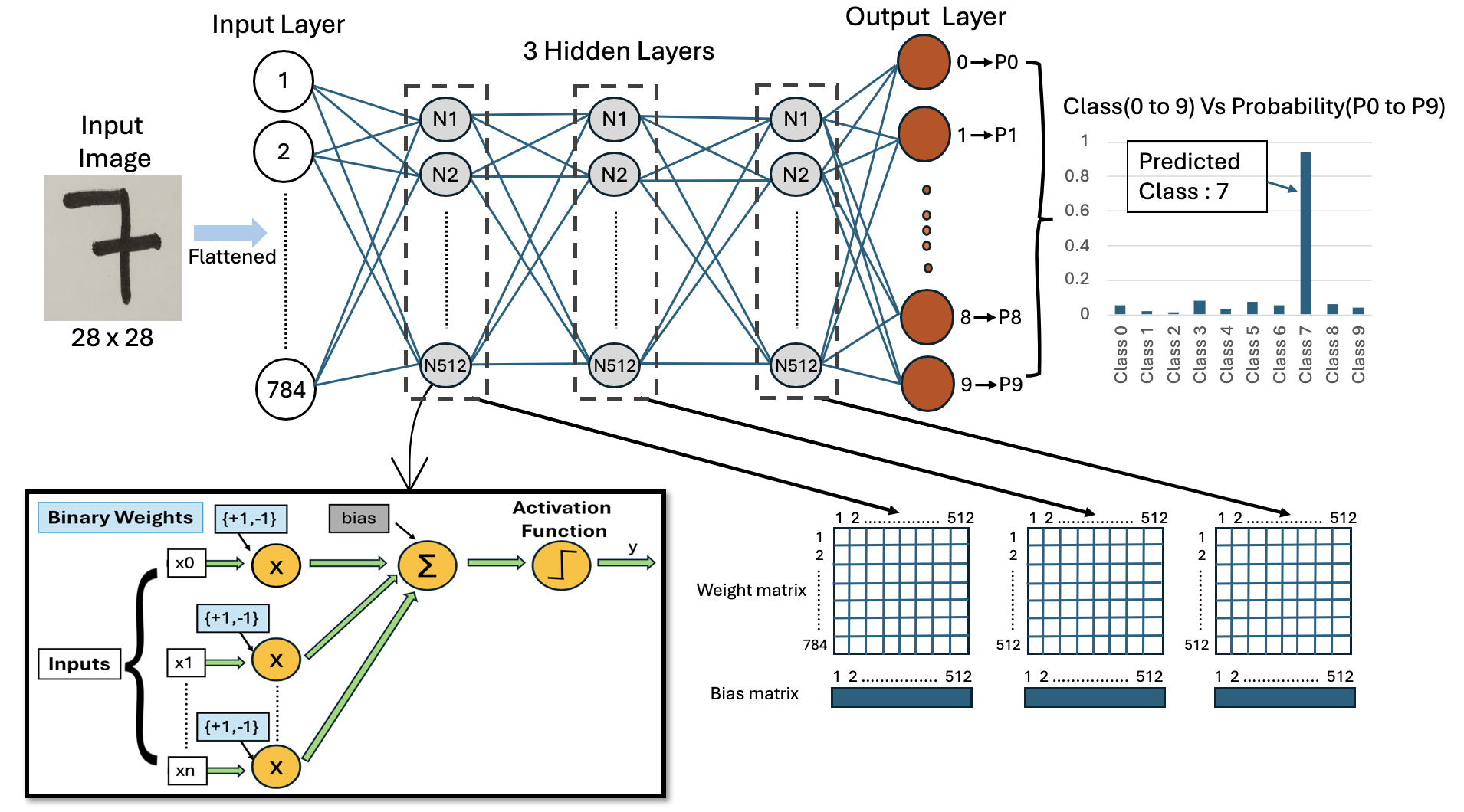}  
    \caption{BNN model used for digit classification.}
    \label{fig:PUF_binary}
\end{figure*}

To overcome this limitation, researchers have introduced Binarized Neural Networks (BNNs)~\cite{courbariaux2016bnn} as an efficient alternative. In BNNs, both the weights and the activations are restricted to binary values, typically +1 and -1, instead of full-precision floating point numbers. This simplification brings significant benefits as follows:
reduces model size, eliminates the need for costly multiplications instead uses efficient bitwise logic,
reduces power consumption and increases inference speed, making BNNs ideal for edge computing scenarios. See Figure \ref{fig:PUF_binary} for an illustration of BNNs.

Because of these advantages, BNNs are gaining popularity in applications such as smart sensors~\cite{smartsensor}, real-time vision systems~\cite{realtime}, and low-power AI accelerators~\cite{andri2021chewbaccann}. To further optimize BNN execution, many hardware implementations~\cite{wang2023benchmarking} utilize Resistive Random Access Memory (RRAM) devices arranged in crossbar arrays. These arrays serve as both storage and computation units: model weights are stored as resistive states, and inference is performed using analog or in-memory computation. This design minimizes the need to transfer data between memory and processor, leading to faster and more energy-efficient inference.

However, these same benefits introduce serious security concerns. Since the weights are stored permanently in non-volatile memory, they remain physically accessible after deployment. An attacker with physical access to the hardware can extract the model parameters using invasive techniques such as Micro-probing~\cite{microprobing}, Energy Dispersive X-ray Spectroscopy~\cite{edx} and Laser-based fault injection~\cite{laser}, optical imaging~\cite{opticalimaging}, etc.
Such attacks may compromise the intellectual property of the model provider and may also expose private or sensitive data encoded in the weights.

To address these threats, various protection mechanisms have been explored. A promising approach is the use of Physically Unclonable Functions (PUFs) to secure model parameters at the hardware level. A PUF-based circuit can generate a unique and repeatable binary key based on manufacturing variations that are practically impossible to replicate. This response serves as a device-specific cryptographic key. By binding a BNN model to a device’s PUF key, the model can only be decrypted and executed on the device it was intended for, ensuring protection against cloning, theft, or unauthorized reuse~\cite{dorfmeister2024puf,zheng2022device}.

Recently, a PUF-based protection scheme was proposed by Rajendran et al. in \cite{rajendran2025securing}, wherein the PUF key is used to transform both the weight and bias matrices and Batch Normalization (BN) parameters of a 3-layer BNN during deployment (see again Figure \ref{fig:PUF_binary}). The same PUF key is then regenerated at inference time to decrypt the model just before execution. Their method includes two \textit{transformation} strategies for the binary weight matrices of 512 columns :
\begin{enumerate}
    \item \textbf{Inversion-based transformation:} columns of the weight matrices are multiplied by $-1$ if the corresponding bit is $1$ in the 512-bit PUF key.
    \item \textbf{Swapping-based transformation:} adjacent columns in the matrices are conditionally swapped based on the 256-bit PUF key, and a similar transformation is applied to the BN parameters to maintain consistency.
\end{enumerate}

Since this work focusses on the column swapping scheme for the security of BNN model, we recollect it in further detail:
\begin{itemize}
    \item A $256$-bit PUF key is generated on the device.
    \item For each bit of the key, pairs of adjacent columns in the weight matrix (of size $512 \times 512$) 
    are swapped. If the first key bit is $1$, then the first two columns are swapped; else that pair of columns is not swapped. This procedure is repeated for the remaining 255 pairs of columns.
    \item The same column-swapping pattern is applied to the $\gamma$ (scale) and $\beta$ (shift) vectors in the BN layers.
    \item The transformed model is mapped to the RRAM crossbar for deployment.
    \item At inference time, if the correct PUF key is available (i.e. the model runs on its intended device), the same key is regenerated and used to reverse the column swapping, thereby restoring the original model parameters. 
\end{itemize}
This process is illustrated conceptually in Figure
~\ref{fig:swapping_matrix},
where the model owner transform (i.e.,``encrypts'') the model using the PUF key and ``decrypts'' it before inference using the same response.

However, if the model is run on a different device or without proper decryption, the weights and BN parameters remain in their transformed state. In \cite{rajendran2025securing} it is reported that under such conditions, model accuracy for classification on the MNIST database drops from 96.04\% (original) to 52.96\% (encrypted), thereby rendering the encrypted model unusable. Although originally demonstrated on RRAM-based crossbar architectures, our recovery method operates using accuracy feedback. It can therefore be applied to any IoT device that deploys a PUF-encrypted BNN model, regardless of the underlying memory technology.

\subsection*{Our Contribution}
This paper investigates the possibility of attacking the column-based swapping transformation to protect the BNN model parameters. Specifically, we show that we can efficiently recover the PUF key, and hence, the BNN model parameters, from the transformed/encrypted binary weight matrices.

We propose a differential cryptanalysis-based key recovery method that operates directly on the encrypted model without any prior knowledge of the PUF key. We recover the PUF key bit-by-bit by observing the change in classification accuracy averaged over many classifications when guessing the value of that particular bit. Recall that if the corresponding key bit is 1, then the columns would have been swapped, else it would not have been swapped. Since do not know the value of the key bit, we will enumerate both the possibilities for that bit and apply the reverse transform. Our hypothesis is that for the right guess of the key bit, we will obtain a slight improvement in the accuracy. This way we can recover the key bit-by-bit. The complexity of our attack is linear in the size of the PUF key unlike the brute-force attack that has exponential complexity. We stress that our attack does not assume any prior knowledge of the PUF key nor do we attack the PUF circuit itself.

We evaluate our method on a BNN trained on the MNIST dataset, using the same architecture and encryption process as described in the original scheme. Evaluated across 100 encrypted model variants, our method could recover 85\% of the PUF key (see Table~\ref{tab:summary_metrics}), and recover the
BNN model up to 93\% classification accuracy compared to the
original model’s 96\% accuracy. Our attack is very efficient and it takes a couple of minutes to recovery the PUF key and the model parameters.

Section~\ref{sec:background} summarizes the scheme from \cite{rajendran2025securing}, Section~\ref{sec:methodology} details our proposed methodolgy/attack, Section~\ref{sec:results} describes the experimental setup, optimisation and results, and Section~\ref{sec:conclusion} concludes with future directions.

\section{Recap of Encryption Scheme from \cite{rajendran2025securing}}
\label{sec:background}

In their work on securing BNN models deployed on memristive crossbars,~\cite{rajendran2025securing} propose a novel encryption method based on column-swapping guided by a PUF key. Their architecture consists of a BNN trained on the MNIST dataset, comprising three fully connected (FC) layers of size $512 \times 512$ and a final output layer with 10 neurons (see Figure~\ref{fig:PUF_binary}). Each layer contains binarized weights $W^b_{j,k} \in \{-1, +1\}$ and Batch Normalization (BN) parameters $\gamma_k$ and $\beta_k$. To simplify hardware implementation, the BN layers are integrated into the inference logic as thresholding operations. 

\begin{figure*}[!htbp]
    \centering
    \includegraphics[width=0.99\linewidth]{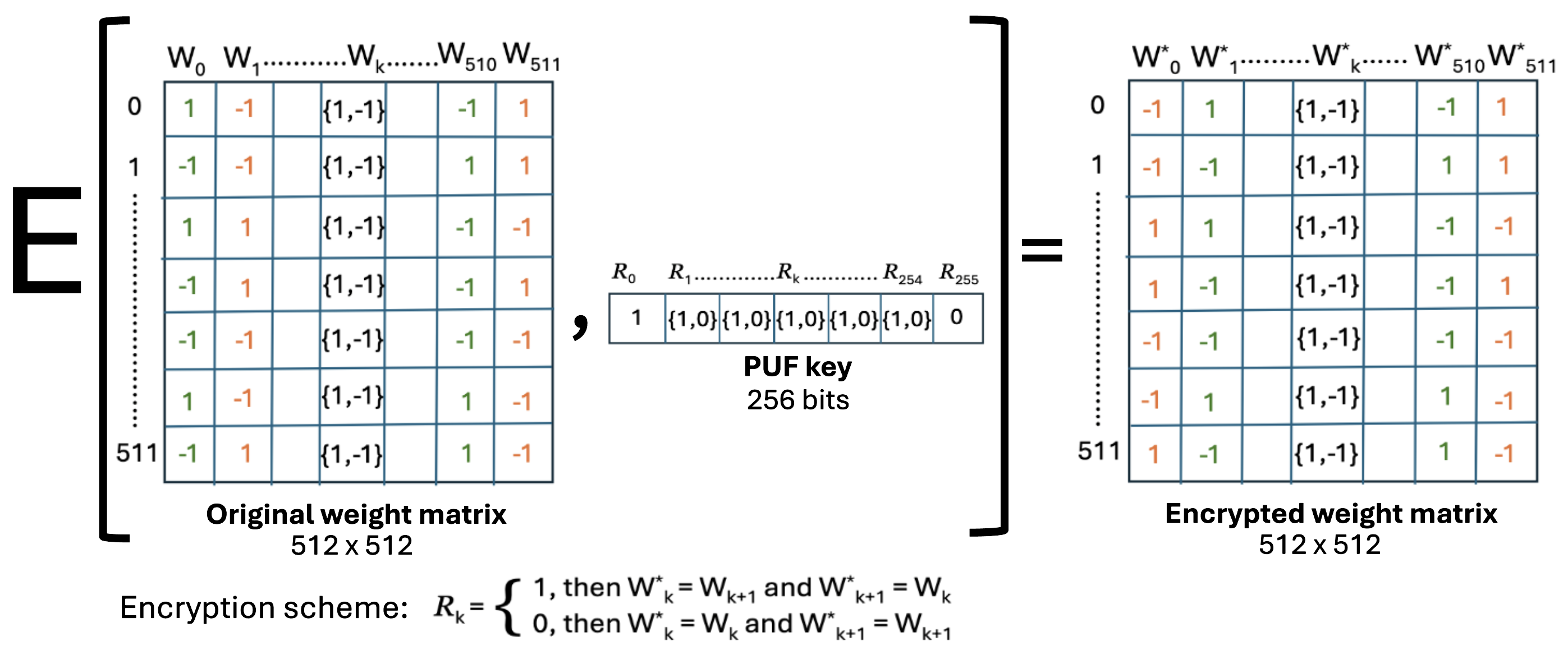}  
    \caption{Proposed swapping scheme.}
    \label{fig:swapping_matrix}
\end{figure*}

To protect this model from physical readout attacks on non-volatile RRAM crossbars, the authors propose applying a transformation to the binary weight matrices that are of dimension $512 \times 512$, and BN parameters(512 size vector) based on a 256-bit PUF response for every two columns. Their column swapping scheme operates as follows: Let the binary PUF response be $\mathbf{R} = [R_0, R_1, \dots, R_{255}]$, where each bit $R_k \in \{0, 1\}$ determines whether a swap should occur between columns $2k$ and $2k+1$ of the weight matrix and corresponding BN parameters. The exact column swapping operations (for $R_k = 0$ or $1$) are shown in Figure~\ref{fig:swapping_matrix}.

After applying this transformation, the encrypted model is programmed into the RRAM crossbar. At inference time, if the correct PUF response $\mathbf{R}$ is regenerated on the target device, the same swap logic can be applied again to recover the original order of weights and BN parameters. This design ensures that the model yields correct outputs only on authorized hardware with access to the correct PUF key.

If the PUF key is unavailable or incorrect, the model remains encrypted and its accuracy significantly drops.~\cite{rajendran2025securing} reports a degradation of classification accuracy from $96.04\%$ (original unencrypted matrix) to $52.96\%$ accuracy when inference is attempted without performing the correct decryption step.

This protection method offers a high brute-force security strength. Since the PUF response per layer consists of 256 bits, the total security across the three layers is estimated as $2^{3 \times 256} = 2^{768}$, assuming independent keys per layer. We target this scheme, testing its vulnerability to recovery attacks.

\section{Proposed Methodology}
\label{sec:methodology}

\subsection{Threat Model}

We consider a threat model in which the attacker has physical access to the RRAM crossbar. The adversary is assumed to have the following capabilities: (1) the ability to extract the transformed weight matrices \( W^{sb}_{j,k} \) and the corresponding batch normalization (BN) parameters \( B^s_k \) from the deployed Binarized Neural Network (BNN) using techniques such as micro-probing; and (2) full knowledge of the BNN architecture, including all layer configurations and activation functions. However, the key security assumption is that the attacker does not possess the Physical Unclonable Function (PUF) response \( R \), which is used to permute the model parameters during deployment. The attacker's goal is to reconstruct \( R \) and thereby restore the original model.


In line with realistic deployment scenarios such as edge devices or cloud-hosted APIs, we further assume that the adversary does not have access to the original full training dataset. At best, they may obtain a limited set of labeled inputs. To simulate this constraint consistently across all evaluations, we assume the attacker has access to 5,000 labeled samples from the MNIST test set. This upper bound represents a practical yet generous estimation of what could realistically be gathered by an adversary.

We next introduce our proposed key recovery technique, that we call as \textit{single-bit recovery method}. This approach aims to reconstruct the 256-bit PUF key \( R \) by evaluating the impact of individual bit flips on model accuracy. 

\subsection{Single‑bit Recovery Method}

The single‑bit recovery method reconstructs the 256‑bit PUF key $\mathbf R$ by iteratively testing, for each bit position $k$, whether swapping that adjacent column pair improves classification accuracy. Given test data $(X, y)$, where $X$ denotes the inputs and $y$ the corresponding labels, we begin from a guess whose bits are all zero $0^{256}$, we flip one bit at a time, evaluating model accuracy with $R_k=0$ versus $R_k=1$ on a fixed validation set. We then retain the value that yields higher accuracy averaged over many classifications.  This process is repeated over several passes to refine earlier decisions and mitigate inter‑bit dependencies, ultimately producing a recovered key that maximizes the model’s performance (see Algorithm~\ref{alg_single_bit_recovery}).

\begin{algorithm*}[!htbp]
\caption{Single-bit Recovery Method for Encrypted BNN Model}
\label{alg_single_bit_recovery}
\begin{algorithmic}[1]
\State \textbf{Input:} Encrypted three-layer BNN model $\mathcal{M}$, test data $(X, y)$
\State \textbf{Output:} $R_{\text{recovered}} \in \{0,1\}^{256}$
\Function{RecoverPUF}{$\mathcal{M}, X, y$}
    \State Load model and extract quantized layer weights and batch normalization parameters.
    \State Initialize $R_{\text{recovered}} \gets \mathbf{0}^{256}$.
    \For{$k = 0$ to $255$}
        \State Evaluate model accuracy using $R_k = 0$ and again using $R_k = 1$, following the swapping protocol described.
        \State Select the value of $R_k$ that yields higher model accuracy.
        \State $R_{\text{recovered}}[k] \gets R_k$
    \EndFor
    \State \Return $R_{\text{recovered}}$
\EndFunction
\end{algorithmic}
\end{algorithm*}

\begin{algorithm*}[!htbp]
\caption{Block-based Recovery Method for Encrypted BNN Model}
\label{alg_block_based_recovery}
\begin{algorithmic}[1]
\State \textbf{Input:} Encrypted three-layer BNN model $\mathcal{M}$, test data $(X, y)$, block size $G$
\State \textbf{Output:} $R_{\text{recovered}} \in \{0,1\}^{256}$
\Function{RecoverPUF}{$\mathcal{M}, X, y, G$}
    \State Load model and extract quantized layer weights and batch normalization parameters.
    \State Initialize $R_{\text{recovered}} \gets \mathbf{0}^{256}$.
    \For{each block $\vec{b}$ of $G$ consecutive bits in $R_{\text{recovered}}$}
        \State For all $2^G$ possible combinations for block $\vec{b}$, set $R_{\vec{b}}$ accordingly and evaluate model accuracy.
        \State Select the block pattern yielding the highest accuracy.
        \State Set $R_{\vec{b}}$ in $R_{\text{recovered}}$ to this block pattern.
    \EndFor
    \State \Return $R_{\text{recovered}}$
\EndFunction
\end{algorithmic}
\end{algorithm*}

However, this approach has some limitations: (1) correlation between a bit and the change in classification accuracy is usually not strong compared to a block of bits, 
and (2) the method is \emph{greedy}, meaning that once a bit is set based on local accuracy during a pass, it is not revisited, potentially leading to suboptimal recovery of the key.

\subsection{Block-based Recovery Method}

To overcome the above limitations, we extend this idea to a block-based recovery method. Instead of recovering one bit of the PUF key at a time, we try to recover a block of $G$ many bits at a time. Given test data $(X, y)$, where $X$ denotes the inputs and $y$ the corresponding labels, we determine for every block the right combination of $G$ many bits that maximizes the average classification accuracy. This implies that the adversary must now enumerate $2^G$ possible combinations. 
This captures inter-bit dependencies and significantly improves the accuracy of key recovery with a manageable increase in computational complexity. 

After evaluating all possible patterns for a given block, the pattern that yields the highest validation accuracy is selected and fixed for that block. This process is repeated sequentially over all blocks, progressively building the estimate of the recovered key $R$ (see Algorithm~\ref{alg_block_based_recovery}). 

While the block-based method requires $2^G$ model evaluations per block, the total computational cost amounts to $2^G \times (256 / G)$, where $G$ is the group size. This results in a higher complexity compared to the single-bit approach, but generally enables superior key recovery, particularly when significant correlations exist between bits within the same block. In practice, choosing an appropriate block size $G$ is essential for trading off computational overhead and recovery accuracy, thus facilitating efficient and robust reconstruction of the PUF key.

\subsection{Parallelized Inference for Efficient Key Recovery}

To further improve the runtime efficiency of recovering multiple PUF keys from transformed BNN models, we parallelize the model inference involved in key recovery. Instead of evaluating each candidate pattern for a group sequentially, we distribute these inference tasks across available CPU cores. Each core independently tests different candidate patterns on the test dataset at the same time, fully utilizing modern multi-core processors. Since these pattern evaluations are independent, this approach makes it possible to check many possible solutions simultaneously, significantly accelerating the recovery process.
This parallelization strategy achieves a substantial reduction in overall wall-clock time required for recovery and offers near-linear scalability with the number of processor cores. By maximizing core utilization during the inference stage, we ensure that each group’s candidate space is explored efficiently. Independent evaluation also improves robustness, as any issue encountered during the inference of one candidate does not affect the processing of others, making high-throughput key recovery both practical and reliable.

\section{Results and Discussion}
\label{sec:results}
We replicated the BNN architecture from~\cite{rajendran2025securing}, which consists of three fully connected (FC) layers with 512 neurons each, followed by a 10-neuron output layer. The network was trained on the MNIST dataset. The dataset consists of 60000  and 10000, training and test images, respecitvely.

\subsection{Experimental Setup}
To implement this architecture, we used Larq~\cite{geiger2020larq}, an open-source deep learning library built on TensorFlow that is specifically designed for BNNs. Larq provides quantized layers, binary-compatible constraints, and training tools optimized for efficient BNN development and deployment.

We trained the BNN for 10 epochs with a batch size of 128, achieving an accuracy of 96.45\%. We then applied the column swapping encryption scheme described in~\cite{rajendran2025securing} to the FC1, FC2, and FC3 layers. For each encrypted variant, we generated an independent 256-bit random binary vector, simulating the PUF-derived keys for the three layers. For ease of experiments, we used the same PUF key for all three layers. These keys were then used to apply column swaps in each corresponding weight matrix and its associated batch normalization parameters. By repeating this process with different key sets, we constructed a total of 100 encrypted model variants, each representing a unique key-protected version of the original BNN.

To evaluate the robustness of this scheme, we implemented our recovery algorithm. For each variant, we applied our method using block sizes of 4 and 8, recovering the unknown PUF key $R$ layer by layer. Evaluation was conducted on a 5{,}000-image validation subset from the MNIST test set.

We used the following metrics to assess performance:
\begin{itemize}
    \item \textbf{Key Bit Recovery Accuracy}: Percentage of correctly recovered bits in the PUF key $R$. This is measured by the Hamming distance between the original key and the recovered key expressed. 
    For example, if the original key is $1010$ and the recovered key is $1000$, then we define the key bit recovery accuracy as $75\%$. 
    \item \textbf{Recovered Model Accuracy}: The model's classification accuracy after recovery.
\end{itemize}

Experiments were conducted on a virtualized environment hosted on a VMware Virtual Platform. The system was equipped with 10 × Intel(R) Xeon(R) Silver 4116 CPUs @ 2.10 GHz and 16GB RAM (11.6GB available), running a Linux OS with kernel-level monitoring via cockpit. The source code we have developed for the experiments is available at \url{https://github.com/bijeet1221/Attack-on-a-PUF-based-Secure-Binary-Neural-Network}.

\subsection{Analysis}

We evaluated the effectiveness of our block-based recovery method on 100 transformed variants of the BNN. The evaluation compares the recovery performance for two block sizes: 4 and 8. 
Comprehensive aggregated results across all 100 variants are reported in Table~\ref{tab:summary_metrics}, highlighting average recovered  model accuracy and average recovered PUF key bit match accuracy for both block sizes. This is compared against the original model's accuracy as well when it is encrypted.

\begin{table*}[!htbp]
\caption{Recovery performance metrics for different block sizes. }
\label{tab:summary_metrics}
\centering
\begin{tabular}{@{} lccc @{}}
\toprule
\textbf{Metric} & \textbf{Single Bit} & \textbf{Block Size 4} & \textbf{Block Size 8} \\
\midrule
Original Model Accuracy                & 96.45\%               & 96.45\%           & 96.45\%           \\
Encrypted Model Accuracy               & 57.99\%               & 57.99\%            & 54.05\%            \\
Recovered Model Accuracy               & 93.47\%               & 93.63\%           & 93.92\%           \\
Recovered Bit Match Accuracy           & 85.14\% (217/256)               & 85.63\% (219/256) & 86.99\% (222/256) \\
Average Recovery Time           & 7.26~min               & 14.5~min & 59.27~min \\
\bottomrule
 \footnotesize \textit{Note:} Results for Block Size 8 are for 10 PUF Key Variants.

\end{tabular}

\end{table*}

Our proposed method demonstrates effective recovery of the PUF key, with block size 8 outperforming block size 4 in both key bit recovery accuracy and restored model accuracy. Specifically, key bit recovery accuracy improves from 85.63\% (block size 4) to 86.99\% (block size 8), while the average recovered  model classification accuracy increases from 93.47\% to 93.92\%.

This improvement suggests that larger blocks capture inter-bit dependencies more effectively, resulting in more accurate reconstructions. However, this comes at the cost of increased computational complexity, 256 patterns must be evaluated per block of size 8, compared to only 16 for block size 4.

Despite the higher complexity, both configurations yield restored accuracies that approach the original untransformed model's performance (96.45\%). 

\subsection{ Comparison with Naive Reverse Engineering}

To further assess the effectiveness of our block-based recovery method, we compare it against a reverse engineering baseline. In this alternative approach, an adversary attempts to reconstruct the original model by training a BNN from scratch using test data only. The adversary in this reverse engineering setup operates under the same assumptions outlined in the threat model of the proposed methodology. We assume the attacker has access to 5,000 labeled samples from the MNIST test set. We use the same size as that of our block-based recovery method to maintain consistency in experimental conditions and ensure a fair comparison between the two approaches.

The reverse-engineered model achieves an accuracy of 59\%. In comparison, our block-based recovery method attains a significantly higher accuracy of 93\%, without requiring model retraining or gradient-based optimization. These results highlight the efficiency and effectiveness of our approach in recovering the original model structure, demonstrating the vulnerability of the column-swapping encryption scheme to structural analysis and white-box attacks.

\section{Conclusion}
\label{sec:conclusion}

We demonstrated that the PUF-based column swapping encryption proposed by~\cite{rajendran2025securing} is susceptible to our block-based recovery attack, recovering the PUF key with over 85\% accuracy and restoring model performance to within 2\% of the original. 
Our attack exposes critical vulnerabilities in the column-swapping encryption scheme in~\cite{rajendran2025securing}. By reducing the effective search space from $2^{3 \times 256}$ to a manageable number of evaluations, the proposed block-wise method significantly weakens the security guarantees of the original design. Although increasing block size improves recovery, it also raises computational demands, requiring a careful balance between attack efficiency and resource constraints.
Our attack necessitates stronger protections for BNNs.
 
Future work could explore improving the efficacy of our attack by extending to larger block sizes, and extending our attack to the inversion method also proposed in ~\cite{rajendran2025securing}. We could also look into developing countermeasures to enhance resilience against our attacks. To support reproducibility and foster further research, we have made our implementation code publicly available.

\section*{Acknowledgments}
This work was partially funded by the Privacy-Preserving Data Processing and Exchange for Sensitive Data in the National Digital Public Infrastructure (P3DX) project grant for Srinivas Vivek.

\bibliographystyle{IEEEtran}
\bibliography{myContents/references}

\newpage

\end{document}